\documentclass[aps,prb,twocolumn,groupedaddress,showpacs]{revtex4-1}
\usepackage{amsmath}
\usepackage{amsfonts}
\usepackage{epsfig}
\usepackage{subfigure}
\usepackage{color}
\usepackage{graphicx}
\begin{document}

\title{Optimized Confinement of Fermions in Two Dimensions}

\author{J.D. Cone,$^1$ S. Chiesa,$^2$ V.R. Rousseau,$^3$ G.G.
Batrouni,$^4$ and R.T. Scalettar$^1$}

\affiliation{$^1$Physics Department, University of California, Davis,
California 95616, USA}

\affiliation{$^2$Department of Physics, College of William and Mary,
Williamsburg, VA 23185}

\affiliation{$^3$Department of Physics and Astronomy,
Louisiana State University,
Baton Rouge, LA 70803}

\affiliation{$^4$INLN,
Universit\'e de Nice-Sophia and Institut Universitaire de France, CNRS;
1361 route des Lucioles, 06560 Valbonne, France}

\begin{abstract}
One of the challenging features of studying model Hamiltonians with cold
atoms in optical lattices is the presence of spatial inhomogeneities
induced by the confining potential, which results in the coexistence of
different phases. This paper presents Quantum Monte Carlo results comparing 
methods for confining fermions in two dimensions, including conventional
diagonal confinement (DC), a recently proposed `off-diagonal
confinement' (ODC), as well as a trap which produces uniform density in the lattice.  At constant entropy and for currently accessible temperatures, we show that the current DC method results in the strongest magnetic signature, primarily because of its judicious use of entropy sinks at the lattice edge. For $d$-wave pairing, we show that a constant density trap has the more robust signal and that ODC can implement a constant density profile.
This feature is important to any prospective search for superconductivity in optical lattices.
\end{abstract}

\maketitle

\noindent
\underbar{Introduction:}
Optical lattice emulators (OLE) control ultra-cold atomic gases with 
lasers and magnetic fields to create experimental realizations of 
quantum lattice models of bosonic or
fermionic particles.  For bosons, classic signatures of low temperature
correlated states- superfluidity and the Mott transition- have been
explored now for a decade \cite{greiner02}.  For fermions, quantum degeneracy
has been established through the observation of a Fermi surface
\cite{kohl05}, as has the Mott transition \cite{jordens08,schneider08}.  The
observation of magnetic order is the next immediate experimental
objective \cite{trotzky08,bloch08,jo09,greif10}.  One ultimate goal is
resolving the long-standing question of whether the doped 2D fermion
Hubbard Hamiltonian has long-range $d$-wave superconducting
order\cite{scalapino}.

Optical lattice experiments face at least two major obstacles in
simulating the fermion Hubbard model.  The first is achieving low enough
temperature to pass through phase transitions and into reduced
entropy ordered phases.  
Present limits in experiments are to
temperatures $T\sim t$ (the near-neighbor hopping energy), and to local entropies per atom 
$\sim 0.77k_{B}$ \cite{jordens10}, values which are at the border for observing
short-range magnetic order.


The other obstacle, which we will be addressing in this paper, is 
inhomogeneity arising from the confining potential
\cite{batrouni02}.  The external field conventionally used to trap cold
atoms in the lattice, a spatially dependent chemical potential which we refer to as `diagonal
confinement' (DC), causes variations in the density per site
$\rho_{\bf i}$, with more atoms, on average, in the center of the lattice
and fewer at the edges. Density plays a
key role in determining which correlations are dominant in interacting
quantum systems, but this is especially true of the fermion Hubbard
Hamiltonian in two dimensions where the magnetic response is very
sharply peaked\cite{vanhove} near half-filling ($\rho=1$).  Various
analytic and numerical calculations suggest that pairing order
also has a fairly sharp optimal filling, $\rho \approx 0.80- 0.85$.

As a consequence of the inhomogeneous density arising from DC, a trapped
gas in an optical lattice will exhibit coexistence of different phases,
complicating the analysis and, potentially, significantly weakening and
blurring the signal of any phase transitions.  To some extent, this loss
of signal is reduced for antiferromagnetism (AFM),
since the Mott gap in a DC trap can produce a
fairly broad region of half-filling,
where AFM is dominant.  But
the problem of observing pairing order with DC seems especially acute
since there is no such protection of the optimal density for
superconductivity.

A recent proposal\cite{rousseau10} to use a reduction to zero of the
hopping at the lattice edge to confine the atoms allows the realization
of systems with more uniform density.  
Such control of hopping parameters is experimentally possible through 
holographic masks\cite{greiner} and is referred to as `Off Diagonal Confinement' (ODC).
Since ODC preserves the particle-hole symmetry of the 
unconfined Hubbard Hamiltonian, this trapping geometry can lead to a 
uniform $\rho=1$ density, and at low entropy, can produce a pure antiferromagnetic phase.

What is unclear is whether ODC is an effectively superior
way to confine fermions in OLE. This is a non-trivial question since OLE 
experiments do not have direct control over the 
temperature- instead, the lattice and trapping potentials are introduced adibatically, so 
optical confinement methods must be compared at fixed entropy. 
 Here we present Determinant Quantum Monte Carlo (DQMC)
\cite{blankenbecler81} calculations which compare systems with ODC and DC
traps, as well as a proposed melding of these confinement methods to create
a constant density (CD) trap. We evaluate the effects of these traps on magnetic
order and $d$-wave pairing correlations across the lattice.

Our key results 
are 1) that the conventional DC trap
yields larger spin correlations than an ODC trap at the 
same, currently accessible entropy, 2) that however, for $d$-wave pairing, a constant density trap has a larger response than possible with DC traps, and 3) that ODC can implement a near constant density profile.

\vskip0.05in
\noindent
\underbar{Trapped Hubbard Model, Computational Methods:}
The Hubbard Hamiltonian in the presence of spatially varying
hopping and chemical potential is,
\begin{eqnarray}
&H& = -\sum_{\langle {\bf ij}\rangle,\sigma}
t_{\bf ij}^{\vphantom{dagger}} \,\,
(c_{{\bf j}\sigma}^{\dagger} c_{{\bf i}\sigma}^{\vphantom{dagger}}
+c_{{\bf i}\sigma}^{\dagger} c_{{\bf j}\sigma}^{\vphantom{dagger}})
\\
&+&U \sum_{\bf i} (n_{{\bf i}\uparrow}-\frac12)
(n_{{\bf i}\downarrow}-\frac12)
- \sum_{\bf i} \, \mu_{\bf i} \,\,
(n_{{\bf i}\uparrow} +n_{{\bf i}\downarrow})
\,\,\,.
\nonumber
\label{hubham}
\end{eqnarray}
Here $c_{\bf i \sigma}^{\dagger}$ ($c^{\vphantom{dagger}}_{{\bf
i}\sigma}$) are creation (destruction) operators for two fermionic
species $\sigma$ on site ${\bf i}$, and $n_{{\bf i}\sigma}$ are the
corresponding number operators.
We will study a 2D square lattice with hopping between pairs of near
neighbor sites $\langle {\bf ij} \rangle$.  For a DC trap, $t_{\bf ij}$ 
is constant and the chemical
potential $\mu_{\bf i}= \mu_{0} - V_{\rm t} (i_x^2 + i_y^2)$ decreases
quadratically towards the lattice edge.  For an ODC trap, instead,
$\mu_{\bf i}$ is constant and the hopping term varies.  Here we choose a
parabolic form $t_{\bf ij} = t_0 -\alpha \, r_{\rm bond}^2$,
where $r_{\rm bond}$ is the distance of the center of
bond $\langle {\bf ij}\rangle$
to the lattice center\cite{rousseau10}.  Particle-hole symmetry for this geometry
implies the density $\rho_{\bf i}=\langle \sum_{\sigma} c_{{\bf i}
\sigma}^{\dagger} c_{{\bf i}\sigma}^{\vphantom{dagger}} \rangle=1$ for
all lattice sites when $\mu_{\bf i} = 0$.

For ODC, we fix energy units by setting
$t_0=1$ at the lattice center.
The parameter $\alpha$, which controls
the hopping decay, is chosen so that $t_{\bf ij} \rightarrow 0$ at the
edge. A similar convention is used for the CD trap, with the addition that now, $V_t$,
as well as $\alpha$ is non-zero.

We characterize and compare traps using nearest-neighbor(nn) magnetic and 
next-nearest-neighbor(nnn) $d$-wave pairing correlation functions:
\begin{eqnarray}
&&S^{+}_{\bf i} =
c^{\dagger}_{{\bf i}\uparrow}
c^{\vphantom{\dagger}}_{{\bf i}\downarrow}
\hskip0.15in
\Delta^{\dagger}_{\bf i} =
c^{\dagger}_{{\bf i}\uparrow}
(c^{\dagger}_{{\bf i}+\hat x \downarrow}
-c^{\dagger}_{{\bf i}+\hat y \downarrow}
+c^{\dagger}_{{\bf i}-\hat x \downarrow}
-c^{\dagger}_{{\bf i}-\hat y \downarrow})
\nonumber \\
&&c_{\rm spin}({\bf i,j}) =
\langle S^{-}_{\bf j} S^{+}_{\bf i} \rangle
\hskip0.5in
c_{\rm pair}({\bf i,j}) =
\langle \Delta^{\vphantom{\dagger}}_{\bf j} \Delta^{\dagger}_{\bf i} \rangle
\label{corrfunc}
\end{eqnarray}

\vskip0.05in
\noindent
\underbar{Use of LDA to simulate traps}:
To simulate different traps, we first compute observables and entropy values
for homogeneous 8x8 lattices using DQMC\cite{blankenbecler81} 
which provides exact results for operator expectation values of the fermion Hubbard 
Hamiltonian\cite{dqmc_lim}.The entropy is obtained via energy integration from $T=\infty$. 

We then use the Local Density Approximation to simulate the 
effects of each trapping method for a much larger lattice. With the LDA, observables for any
position in a trap are determined by the density($\rho$) of the equivalent homogeneous
system- that is, a system  with the same values for $U/t$ , $mu/t$, and $T/t$ as the local  
point in the trapped lattice. So for each trap, we compute these values at each position(as a function  
of r) and determine the spin and $d$-wave pairing correlation values from equivalent homogeneous result 
The accuracy of the LDA for short range correlation functions
has been demonstrated for 2D lattices in the regime of temperature presently 
considered[\onlinecite{chiesa11}].

For a trap at a given temperature, the number of fermions(N) 
and total entropy(S) are obtained by integrating the site
density $\rho(r)$ or entropy per site $s(r)$ across the lattice:
$N = 2\pi \int_{0}^{\infty} r \rho(r) \, dr$  and 
$S = 2\pi \int_{0}^{\infty} r s(r) \, dr$. Discrete lattice sums are not used,
since the simulated lattice sizes(approximately 11,000 sites for $r= 60$) are 
quite large and the functions $\rho(r)$ and $s(r)$ are fitted as continuous curves.

\begin{figure}[t]
\centerline{\epsfig{figure=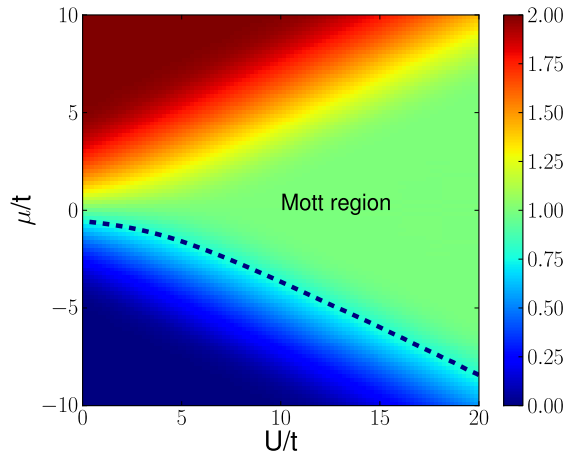,height=6.5cm,width=8.5cm,angle=0,clip}}
\caption{(color online).
The density as a function of $U/t$ and $\mu/t$ for the homogeneous fermion
Hubbard model.  Data were obtained on 8x8 lattices with
$T/t =0.5$. The dotted line is a constant density path ($\rho =.80$) used in the trap comparisons. Using the LDA, density profiles of
inhomogeneous models can be determined by following an appropriate
$(U/t,\mu/t)$ path of local parameters as the lattice position is
changed.
}
\label{rho_contour}
\end{figure}

\begin{figure*}[t]
\includegraphics[scale=0.28]{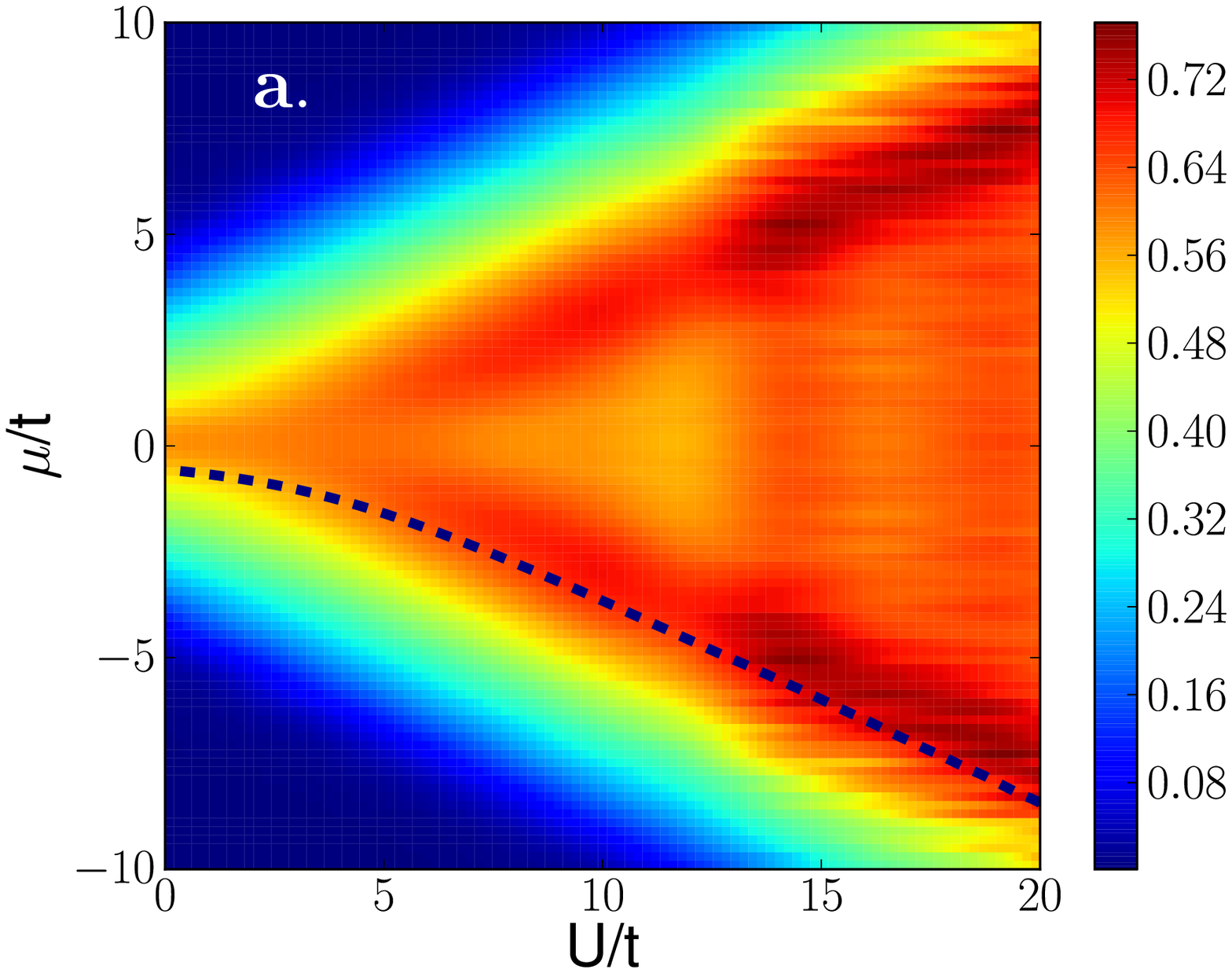}
\includegraphics[scale=0.28]{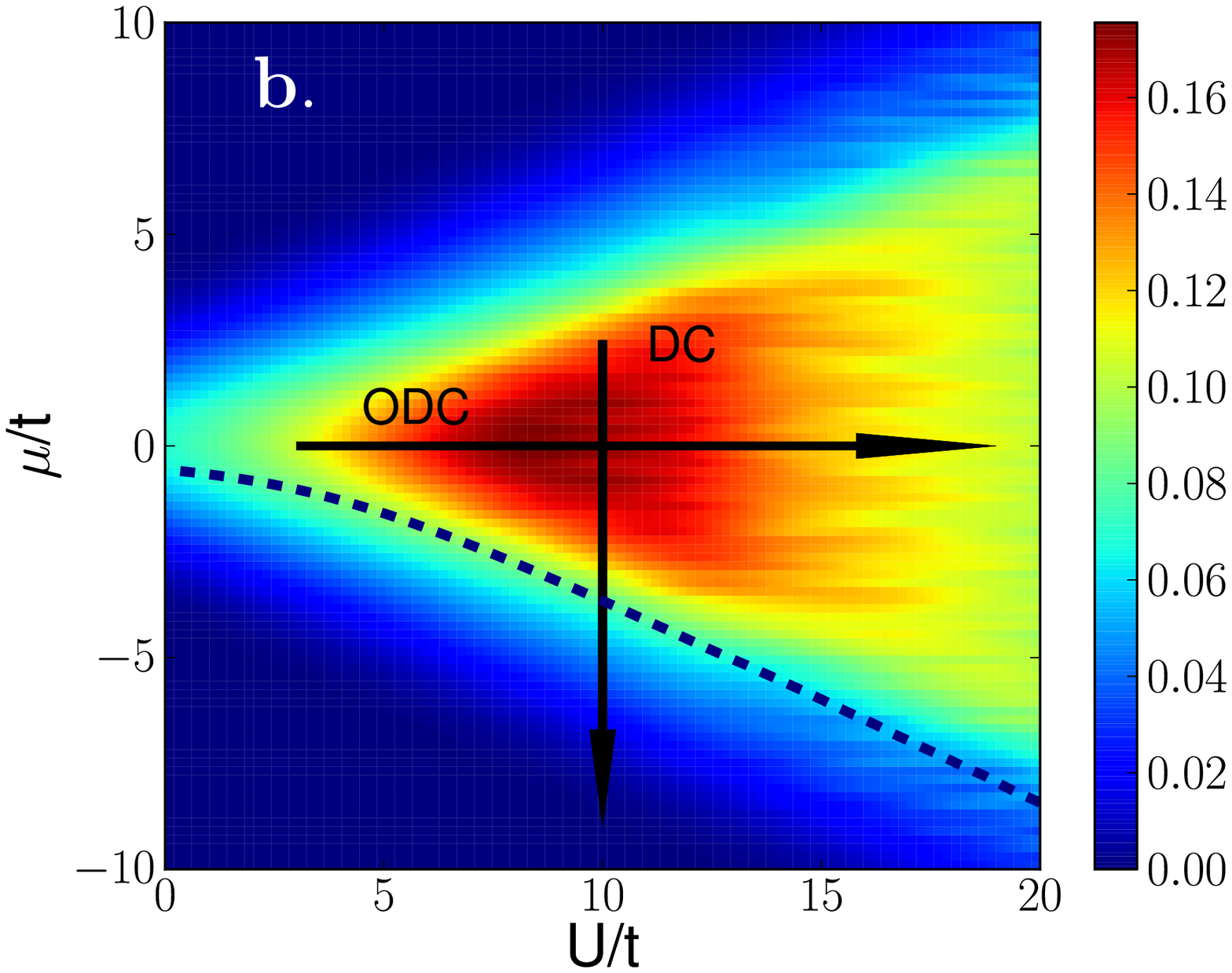}
\includegraphics[scale=0.28]{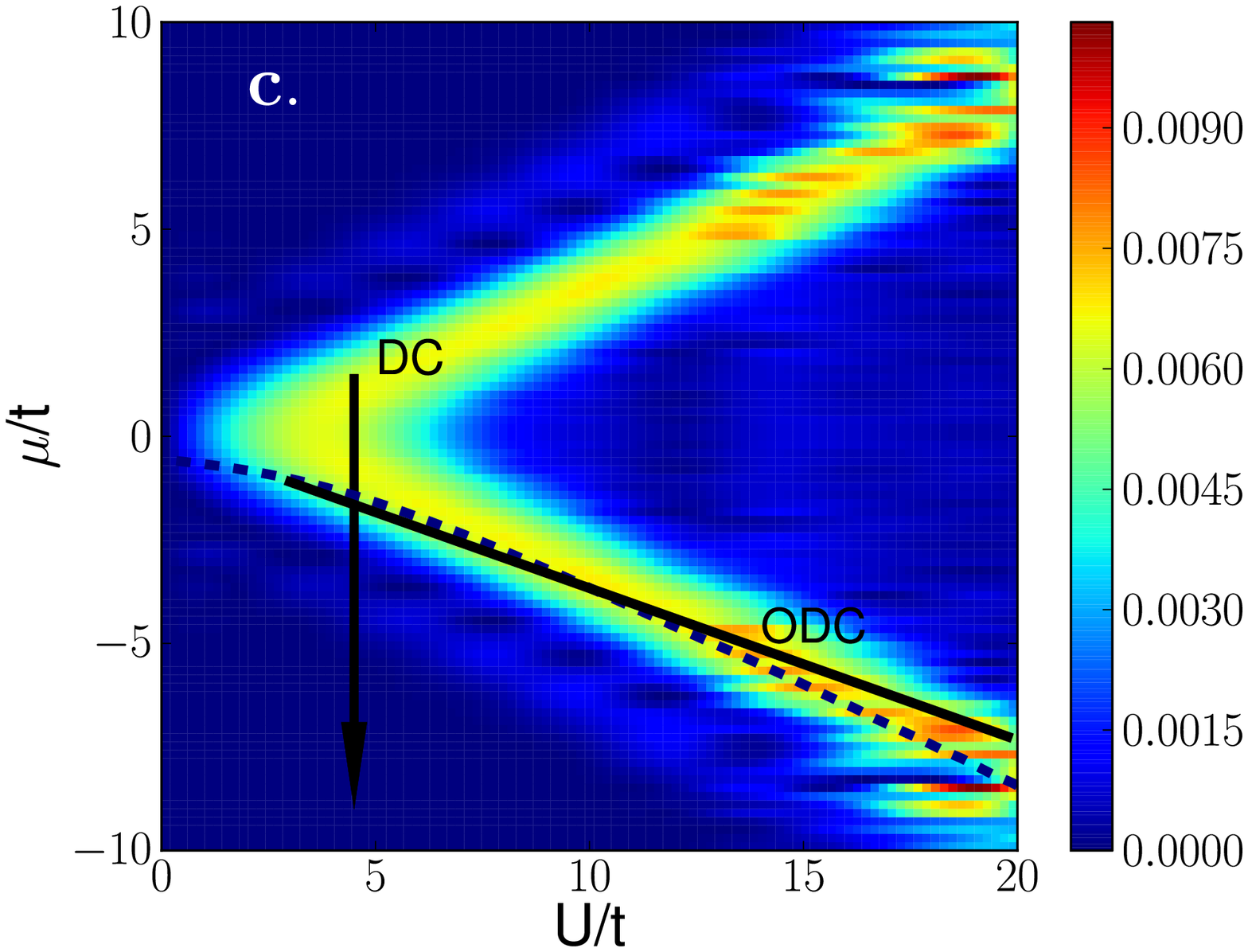}
\caption{(color online).
(a) Entropy per site, (b) spin (nn) correlation, and (c) $d$-wave pairing (nnn) correlation shown as functions 
of $U/t$ and $\mu/t$ for the homogeneous Hubbard Hamiltonian.  Data were obtained on 8x8 lattices with
$T/t =0.5$. Paths for optimal DC and ODC traps are shown as solid lines, with the constant density trap (CD) as a dashed line. The ridge of prominent $d$-wave pairing (c) occurs at $\rho\sim 0.80$ and is nearly linear, so that an ODC path is at close to constant density.
}
\label{all_contour}
\end{figure*}

\vskip0.05in
\noindent
\underbar{Comparing Different Types of Traps:}
Figures ~\ref{rho_contour} and ~\ref{all_contour}a show the variation of the density ($\rho$) and entropy per site with $U/t$ and $\mu/t$ for the uniform Hubbard model at $T/t = .5$. 
Different trapping geometries correspond to the different paths in the $(\mu,U)$
plane. Figures ~\ref{all_contour}b and ~\ref{all_contour}c show the variations in spin(nn) correlation and 
$d$-wave signature, respectively, for the 2D Hubbard model. The arrows in Fig.~\ref{all_contour}b,c indicate the trajectories of sample trap paths projected onto the $U/t - \mu/t$ plane. Note that these figures are at a specific temperature($T/t = 0.5$), while for an actual trap path (ODC for example), T/t will vary across the lattice and we compare traps with fixed entropy, not temperature.

The four parameters $ U/t_0$, $\mu_0/t_0$, $V_t/t_0$, and $\alpha/t_0$ determine the shape and physics for each trap type with the following constraints: DC - $\alpha=0$;
ODC - $V_t=0$; and CD - $V_t$ and $\alpha$ chosen to approximately follow a constant density path.
To estimate the potential pairing and magnetic signature of a trap, we use the following two quantities: (1) the average nnn d-wave pairing correlation and
(2) the average nn spin correlation. Both values are averages per particle over the trap.

We first determined the optimal $U$, $\mu_0$ , $V_t$ and $\alpha$ for each trap type (DC, ODC, CD) by selecting the parameter values yielding traps with the largest
possible pairing or magnetic signature.
These ``optimal'' traps were selected by varying the relevant trap parameters ($U, \mu_0, V_{t}, \alpha$) to maximize a response. Once the optimal parameters were identified by trap type, we proceed to compare traps of different types. In all comparisons, the total number of fermions and total entropy are the same for each trap.

\begin{figure}[t]
\centerline{\epsfig{figure=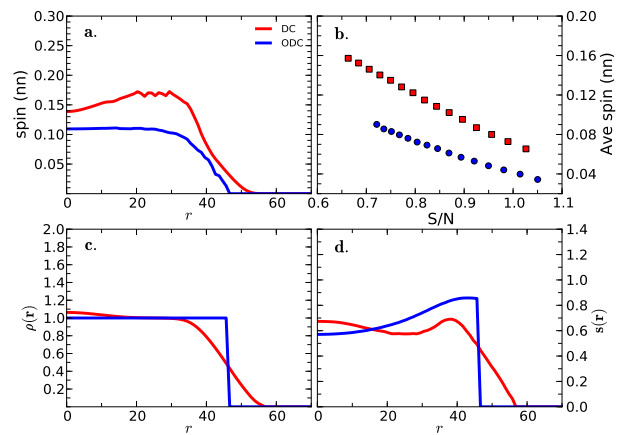,height=6.5cm,width=8.5cm,angle=0,clip}}
\caption{(color online)  $\bf (a)$ Spin (nn) correlation,  $\bf (c)$ density ($\rho$) and $\bf (d)$ entropy per site (s) profiles  are shown as a function of the distance (r) 
from the trap center for two different trap types: DC and ODC, using optimal trap parameters. ODC trap with $\mu$= 0.0 is also a constant density trap ($\rho$ = 1). The average spin(nn) correlation is larger for DC trap (0.14) than ODC (0.08) at the same entropy (0.75) and number of fermions (6600). Panel $\bf (b)$ shows average spin(nn) response as a function of entropy per fermion (S/N) for optimal DC and ODC traps.
}
\label{spin_merge}
\end{figure}

\vskip0.05in
\noindent
\underbar{Magnetic order}:
Figure~\ref{spin_merge} compares an optimal DC trap ($U =10.0, \mu_0 =2.5, V_{t}=0.0039$) with an optimal ODC trap($U =3.0, \mu_0 =0.0, \alpha=0.0004$)  at constant entropy per fermion (S/N = 0.75) and fermion count(6600). As explained in the previous section, these trap parameters were selected to optimize the correlation function- spin(nn) in this case- for fixed N and S. 
Note that the figure panels show only two trap types since, when $\mu_0 = 0.0$, the ODC trap is equivalent to the $\rho = 1$ CD trap. 

We might expect that ODC would lead to a large signal of antiferromagnetism,
since this confinement method allows for a uniform half-filled Mott phase
where magnetic correlations are strong. When an ODC trap is compared with a
DC one at fixed temperature, the former has a stronger magnetic signature.
However, as shown in the figure, the DC trap has a significantly larger average spin
correlation (0.14 vs 0.08) than the ODC trap: an ODC trap is not optimal for
antiferromagnetism when compared to a DC trap at the same entropy.

The reason is that the low density wings in the DC trap
can store a lot of entropy, so that the local entropy nearer to the trap center
is smaller. This central area in the DC trap is effectively at lower temperature
and has higher spin correlations than in an ODC trap where there is
no entropy sink in the wings. Consequences of non-uniform entropy 
distribution have been emphasized previously in [\onlinecite{fuchs2011},\onlinecite{paiva2011}] 

\begin{figure}[t]
\centerline{\epsfig{figure=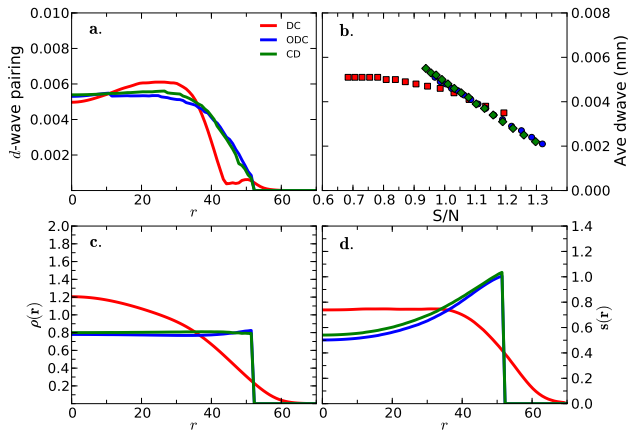,height=6.5cm,width=8.5cm,angle=0,clip}}
\caption{(color online)  $\bf (a)$ $d$-wave pairing correlation,  $\bf (c)$ density($\rho$) and $\bf (d)$ entropy per site (s) profiles are shown as a function of the distance (r) from the trap center for three different trap types: DC, ODC , and constant density (CD) using optimal trap parameters for each type. The average pairing values for CD and ODC traps are 0.0052 and 0.0051, with the DC trap at 0.0046. Entropy per fermion (~0.95) and number of fermions (6600) is the same for each trap. Panel $\bf (b)$ shows average $d$-wave response as a function of entropy per fermion (S/N) for optimal DC, ODC, and CD traps.
}
\label{dwave_merge}
\end{figure}

\vskip0.05in
\noindent
\underbar{Pairing away from Half-Filling:}
We now turn to the question of pairing order. In Fig.~\ref{dwave_merge}, we show results comparing
average next-near-neighbor\cite{whynnn} $d$-wave pairing correlation for optimal DC, ODC, and constant density (CD) traps
at the same entropy per fermion (~0.95) and number of fermions(6600). Optimal parameters obtained for each trap are :
DC ($U =4.5, \mu_0 =1.5, V_{t}=0.00235$), ODC ($U =3.0, \mu_0 = -1.1, \alpha=0.00032$), and CD ($U=3.0, \alpha=0.00032, \rho = 0.80$).

Looking at the trap profiles, we can see
that peak pairing for the DC trap occurs at lattice distances (measured from the center) which correspond to densities between 0.9 and 1.1, but pairing dips outside of this region and falls off rapidly towards the lattice edge. 
The constant density and ODC traps are characterized by larger average pairing values of 0.0052 and 0.0051 (vs 0.0046 for DC). This result can be clearly understood from Fig. ~\ref{all_contour}c by observing how the constant density line ($\rho =0.80$) and ODC path follow the ridge of high $d$-wave pairing response, while the DC trap path cuts through this ridge for only a portion of the trap area. While the pairing difference is not large at this high entropy level, the advantage is expected to grow at lower entropies (see discussion in next section).

Figure~\ref{all_contour}c also emphasizes the narrowness of the optimal $d$-wave response region compared to the wider area 
of magnetic response seen in Fig.~\ref{all_contour}b.  This suggests that OLE trap parameters tuned to follow this ridge of high $d$-wave response will increase the potential for observing superconducting order.

\vskip0.05in
\noindent
\underbar{Summary of trap comparisons:}
Figures~\ref{spin_merge}b and ~\ref{dwave_merge}b summarize the results of our trap comparisons by plotting average spin(nn) and $d$-wave pairing responses against entropy per fermion (S/N). In the range of entropy shown, the DC trap has a larger average spin response than the corresponding ODC trap at the same entropy. For $d$-wave pairing, the DC trap flattens at lower entropy values($S/N \sim 0.6- 0.9$), while the pairing response for the constant density (CD) and ODC traps continues to increase as entropy is lowered. Due to sign problems, we were unable to reach entropy per fermion levels lower than 0.9 for ODC and CD traps, but it is evident from Fig~\ref{dwave_merge}b that the $d$-wave response continues to rise as $S/N$ decreases.

\vskip0.50in
\vskip0.05in
\noindent
\underbar{Conclusions:}
We have evaluated several trapping geometries for fermions in a
2D optical lattice. For magnetic response, the DC trap which is the common experimental technique used in OLE, continues to be the most promising confinement approach, because the DC trap can store excess entropy in its low density wings leaving a low entropy Mott region with large AFM correlations.
We have also shown here that a more robust signal of $d$-wave
pairing is produced with a constant density trap with optimal $\rho\sim 0.80$.
 That is, the local superconducting correlations are large over a significantly greater
fraction of sites. We find that ODC closely follows the constant density line shown in Fig~\ref{all_contour}.

An important conclusion of our work is that while
the search for antiferromagnetic correlations in optical lattices
is aided by the inhomogeneous entropy distribution, 
this is not the case for pairing.  The local entropy is not reduced
in the vicinity of $\rho$=0.80, which is best for d-wave superconductivity.
Thus the same inhomogeneous s(r) which helps the magnetic signal
will weaken the pairing signal.  This is a further argument
for construction of a trap which has constant density.
By providing an optimal confinement template for fermions in two dimensions, we anticipate that the results 
will aid experimenters in determining the physics of the doped Hubbard model.

We acknowledge financial support from ARO Award W911NF0710576 and the DARPA OLE Program, NSF Grant No. OCI-0904972, CNRS-UC Davis EPOCAL LIA joint research grant, and the DOE SciDAC program (DOE-DE-FC0206ER25793). We are grateful for inspiration from G. Thorogood.


\begin{thebibliography}{40}
\bibitem{vanhove}  Bipartite lattices have Fermi surface nesting 
and/or van Hove singularities in the density
of states which magnify the effect of correlations at half-filling.
\bibitem{dqmc_lim}
The effects of discretization of inverse temperature on DQMC 
can be eliminated by extrapolation.
\bibitem{whynnn}
As discussed in Ref.~[\onlinecite{chiesa11}] the nn $d$-wave correlation
function picks up a large contribution from AFM order.  The nnn value is
thus the shortest range case which is a pure measurement of pairing.
\end{thebibliography}
\end{document}